\documentclass[%
reprint,
 amsmath,amssymb,
 aps,
]{revtex4-1}

\usepackage{graphicx}
\usepackage{dcolumn}
\usepackage{bm}

\usepackage{cancel}
\usepackage{xspace}
\usepackage{lineno,hyperref}
\usepackage{xcolor}

\newcommand{\pt}{\ensuremath{p_{\rm T}}\xspace}

\newcommand{\rt}{\ensuremath{R_{\rm T}}\xspace}
\newcommand{\rtmin}{\ensuremath{R_{\rm T}^{\rm min}}\xspace}
\newcommand{\rtmax}{\ensuremath{R_{\rm T}^{\rm max}}\xspace}
\newcommand{\nmpi}{\ensuremath{N_{\rm mpi}}\xspace}

\newcommand{\py}{PYTHIA\xspace}
\newcommand{\hw}{HERWIG\xspace}

\begin{document}

\preprint{APS/123-QED}


\title{Disentangling the hard gluon Bremsstrahlung effects \\ from the relative transverse activity classifier in pp collisions}%
\author{Gyula Benc\'edi}
\author{Antonio Ortiz}
\email{antonio.ortiz@nucleares.unam.mx}
\affiliation{%
Instituto de Ciencias Nucleares, Universidad Nacional Aut\'onoma de M\'exico,\\
 Apartado Postal 70-543, M\'exico Distrito Federal 04510, M\'exico 
}
\author{Antonio Paz}
\affiliation{%
Facultad de Ciencias F\'isico Matem\'aticas,
Universidad Aut\'onoma de Nuevo Le\'on,\\ 
Ciudad Universitaria, San Nicolas de los Garza, Nuevo Le\'on 66450, M\'exico
}

%

\date{\today}

\begin{abstract}

Recently, the so-called relative transverse activity classifier, \rt, has been proposed as a tool to disentangle the particle production originated from the soft and hard  QCD  processes in proton-proton (pp) collisions. \rt is a useful quantity to study particle production in events with exceptionally large or small activity in the transverse region with respect to the event-averaged mean. Contrary to the expectations, the preliminary results of the ALICE Collaboration indicate that, e.g., the proton-to-pion ratio does not exhibit the characteristic enhancement at intermediate \pt in events with large \rt with respect to minimum-bias pp collisions. In this work, we investigate the origin of this effect using the \py~8 and \hw~7 Monte Carlo event generators. The effect is a consequence of a selection bias attributed to wide-angle gluon emissions which creates jets that populate the transverse region. Therefore, we propose a modified version of \rt in order to suppress its sensitivity to hard gluon Bremsstrahlung, and enhance the sensitivity to soft Multiparton Interactions (MPI). This approach could be useful in order to study the particle production in the jet-like region as a function of MPI.  The implementation of these ideas in data will provide more insight into the production mechanisms of hadrons in high-multiplicity pp collisions, and its connection with heavy-ion phenomena. 

\end{abstract}

\maketitle


\section{\label{sec:level1}Introduction}

One of the  most  important  discoveries at the Large Hadron Collider (LHC) and the Relativistic Heavy Ion Collider (RHIC) is the presence of heavy-ion like phenomena, collectivity~\cite{Nagle:2018nvi} and strangeness enhancement~\cite{ALICE:2017jyt}, in high-multiplicity proton-proton (pp) and proton-nucleus collisions. In heavy-ion collisions these observations  are consistent with the formation of a deconfined hot and dense QCD medium, known as the strongly-interacting Quark-Gluon Plasma (sQGP)~\cite{Busza:2018rrf}. However, the presence of sQGP in high-multiplicity pp collisions has not been claimed since no jet-quenching signals have been observed. It is worth mentioning that the search for jet quenching in pp collisions is one of the key measurements which are planned for the future LHC Run 3 by the ALICE Collaboration~\cite{ALICE:2020fuk}.

In order to understand the physics behind high-multiplicity pp collisions from the Monte Carlo (MC) generators perspective~\cite{Sjostrand:2018xcd}, several studies have been reported~\cite{Ortiz:2013yxa,Bierlich:2014xba,Bellm:2019wrh,Bierlich:2020naj}. In particular, we have focused our most recent efforts on the investigation of particle production as a function of the event activity aiming at reducing the selection biases~\cite{Ortiz:2020rwg, Ortiz:2020dph, Ortiz:2021peu}. The selection bias is the main technical issue when one tries to extract the jet-quenching effects from pp data~\cite{Jacobs:2020ptj}. 

When one selects pp collisions with high event activity, the sample is naturally biased towards hard processes. In order to overcome such an effect, it has been proposed explicitly removing the jet contribution from the event activity estimator.  This can be achieved if one identifies an axis which allows for the event-by-event separation of the jet contribution from the underlying event (UE). The underlying event consists of particles that arise from beam-beam remnants and Multiparton Interactions (MPI)~\cite{Field:2012kd}. The direction of the leading charged-particle transverse momentum can be used as a reference axis to build particle correlations with the associated particles in azimuthal angle $\Delta\phi$. And then, the particle production can be studied as a function of the multiplicity in the transverse region ($\pi/3 <|\Delta\phi| <2\pi/3$) quantified in terms of the relative transverse activity classifier \rt. The transverse region is the most sensitive to UE, but it has contributions from initial- and final-state radiation (ISR and FSR) that accompanies the hard scattering. The idea has been originally proposed in Ref.~\cite{Martin:2016igp}, and the properties of \rt reported in Refs.~\cite{Ortiz:2017jaz,Weber:2018ddv}. The preliminary ALICE results~\cite{Tripathy:2021fax} show that for the transverse region, the transverse momentum (\pt) spectra become harder with increasing \rt. At the same time the particle ratios, (anti)proton-to-pion and hyperon-to-pion, do not exhibit a strong dependence on \rt at intermediate \pt ($2-8$\,GeV/$c$)~\cite{Nassirpour:2749160}. This suggests that the data do not follow neither the expected behaviour due to the sensitivity of \rt to MPI~\cite{Ortiz:2013yxa} nor the event activity dependence of particle ratios observed in pp~\cite{Acharya:2018orn}, p--Pb~\cite{Adam:2016dau} and Pb--Pb~\cite{Adam:2015kca} data at the LHC energies. In a previous work,  we have concluded that effects from initial- and final-state radiation are enhanced if one selects events with high activity in the transverse side in pp collisions simulated with PYTHIA~8~\cite{Ortiz:2020dph}. Moreover, we argued that hard Bremsstrahlung gluons can produce an apparent modification of the jet-like yield in events with extremely large    

In this paper, we focus our attention on the UE-dominated transverse region considering a further refinement to distinguish between the more and the less active sides of the transverse region on a per-event basis~\cite{Marchesini:1988hj,Pumplin:1997ix,Acosta:2004wqa}. The purpose of using this geometrical selection criterion is to suppress hard initial and final-state radiation thus increasing the sensitivity of the transverse region to MPI component of the UE. Based on the CDF approach~\cite{Field:2012kd},  the so-called trans-min and trans-max regions are introduced. The trans-min region is insensitive to wide angle emissions from the hard process, while the trans-max region receives contribution from hard initial and/or final-state radiation. The inclusive transverse momentum distributions of charged particles, as well as the proton-to-pion ratio, are studied as a function of multiplicity in the trans-min and trans-max regions. Moreover, we continue our earlier investigation on the di-hadron correlation, and we study the structures of the $\Delta\phi$ distribution for the transverse, trans-max and trans-min  regions. The results are discussed in the context of recent ALICE preliminary results~\cite{Tripathy:2021fax,Nassirpour:2749160}.

The article is organised as follows:  Section II provides information about the \py~8 and \hw~7 Monte Carlo generators which are relevant to the current investigation, discusses the event and particle selection, as well as, the analysis approach. Section III presents the results and discussion, and finally section IV summarizes our results and presents the outlook.

\section{\label{sec:analysis}Monte Carlo generators and Underlying Event observables}

For our analysis, we used \py~8.201~\cite{Sjostrand:2007gs} with the default Monash 2013 tune~\cite{Skands:2014pea} and \hw~7.2~\cite{Bellm:2019zci} with its default tune SoftTune~\cite{Bellm:2019zci} which will be simply labelled as \py~8 and \hw~7 in the following. Both models are fully exclusive hadron-level Monte Carlo generators, containing leading-logarithmic initial- and final-state parton showers, hadronization models and particle decays. 
The focus of the mentioned tunes for these generators are on the description of minimum-bias as well as underlying-event data. It is noteworthy that the SoftTune of \hw~7 provides a calibration of MPI parameters that conforms with an improved model of soft interactions and diffractive processes introduced in \hw~7.1~\cite{HERWIGcollab_a2017a}. With the inclusion of additional contributions, these MC generators are able to describe the underlying-event activity in a hard scattering process, and it was observed earlier that they well describe the measured data by several experiments at the LHC, see e.g.~\cite{Khachatryan:2010pv,Chatrchyan:2011id,Aad:2010fh,ALICE:2011ac}.

In this study, we generated $\sim 3\times10^{9}$   minimum-bias pp events at $\sqrt{s}=5.02$\,TeV  and only final state charged particles were accepted excluding the weak decays of strange particles in order to meet the experimental conditions.
In each event we identified the highest transverse momentum particle (trigger particle) in the mid-pseudorapidity window $|\eta| < 0.8$ corresponding to the acceptance of the main tracking detector of the ALICE experiment~\cite{Aamodt:2008zz}. Only events with a trigger charged-particle in the range $8< p_{\rm T}^{\rm trig.} \leq15$\,GeV/$c$ are considered. The so-called transverse region is the most sensitive to UE, it is defined in terms of  the  relative  azimuthal  angle, $\Delta\phi = \phi^{\rm trig.}-\phi^{\rm assoc.}$, where $\phi^{\rm trig.}$ ($\phi^{\rm assoc.}$) is the azimuthal angle of the trigger (associated) particle. The multiplicity in the transverse region ($N_{\rm ch}^{\rm trans.}$) is obtained considering associated particles with $\pt\geq0.5$\,GeV/$c$ and within $\frac{\pi}{3} < |\Delta\phi| < \frac{2\pi}{3}$. In this analysis, the transverse region is further sub-divided in two regions: $\pi/3<\Delta\phi<2\pi/3$ and $\pi/3<-\Delta\phi<2\pi/3$ which are referred to as transverse I and transverse II.  The overall transverse region corresponds to combining the transverse-I and transverse-II regions. These two distinct regions are characterized in terms of their relative charged-particle multiplicities, $N_{\rm ch}^{\rm trans.\,max}$ and $N_{\rm ch}^{\rm trans.\,min}$, termed trans-max and trans-min. Trans-max (trans-min) refers to the transverse region (I or II) containing the largest (smallest) number of charged particles.

\begin{figure}[t]
\includegraphics[width=0.46\textwidth]{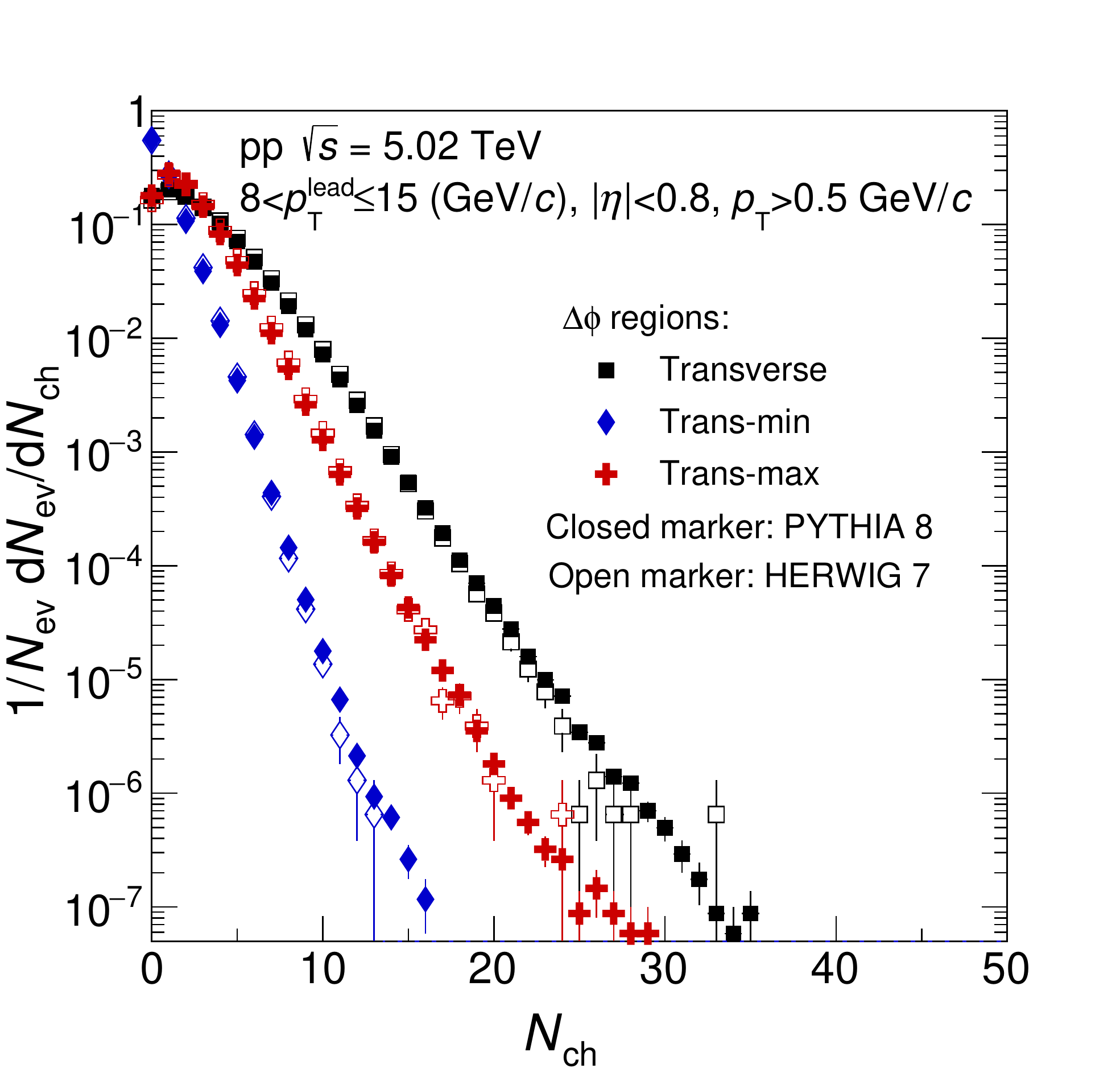}
\caption{Per-event charged-particle multiplicity distributions simulated with \py~8 (full markers) and \hw~7 (empty markers) in pp collisions at $\sqrt{s}=5.02$\,TeV for different topological regions: transverse (squares), trans-min (diamond) and trans-max (cross). The definition of each region can be found in the text.}
\label{fig:fig1}  
\end{figure}

In order to verify our definition of trans-max and trans-min, we extracted the charged particle multiplicity distributions in the kinematic ranges described above. Figure~\ref{fig:fig1} shows the per-event multiplicity distributions for the entire transverse region, and for trans-min and trans-max regions. As expected, the distribution of trans-min has overall the smallest multiplicity reach and average, while the distribution for trans-max has smaller multiplicity reach than that for $N_{\rm ch}^{\rm trans.}$. This feature is also indicated by the average values of the distributions: $\langle N_{\rm ch}\rangle^{\rm trans.\,min}=0.69$, $\langle N_{\rm ch}\rangle^{\rm trans.\,max}=2.02$, and $\langle N_{\rm ch}\rangle^{\rm trans.}=2.71$ for the trans-min, trans-max, and full transverse regions, respectively. The distributions  from \py~8 and ~\hw~7 exhibit the same behaviour, which means that the description of UE is very similar in these models in terms of the charged particle multiplicity.

Experimental data show that the onset of the UE plateau in the transverse region is observed at $p_{\rm T}^{\rm trig.} \sim 8-10$\,GeV/$c$~\cite{Acharya:2019nqn,Adam:2019xpp}, therefore, the mean charged-particle multiplicity in the transverse region has less dependence on $p_{\rm T}^{\rm trig.}$. Thus, it is safe if we classify the events with a trigger particle in the range $8 < p_{\rm T}^{\rm trig.} \leq15$\,GeV/$c$ based on their per-event activity in the transverse region with respect to the mean:

\begin{equation}
R_{\rm T}=\frac{N_{\rm ch}^{\rm trans.}}{\langle N_{\rm ch}^{\rm trans.}\rangle}~.    
\end{equation}

Using $N_{\rm ch}^{\rm trans.\,min}$ and $N_{\rm ch}^{\rm trans.\,max}$, instead of $N_{\rm ch}^{\rm trans.}$, we also can define the quantities $R_{\rm T}^{\rm min}$ and $R_{\rm T}^{\rm max}$, respectively. 

\begin{figure}[t]
\includegraphics[width=0.46\textwidth]{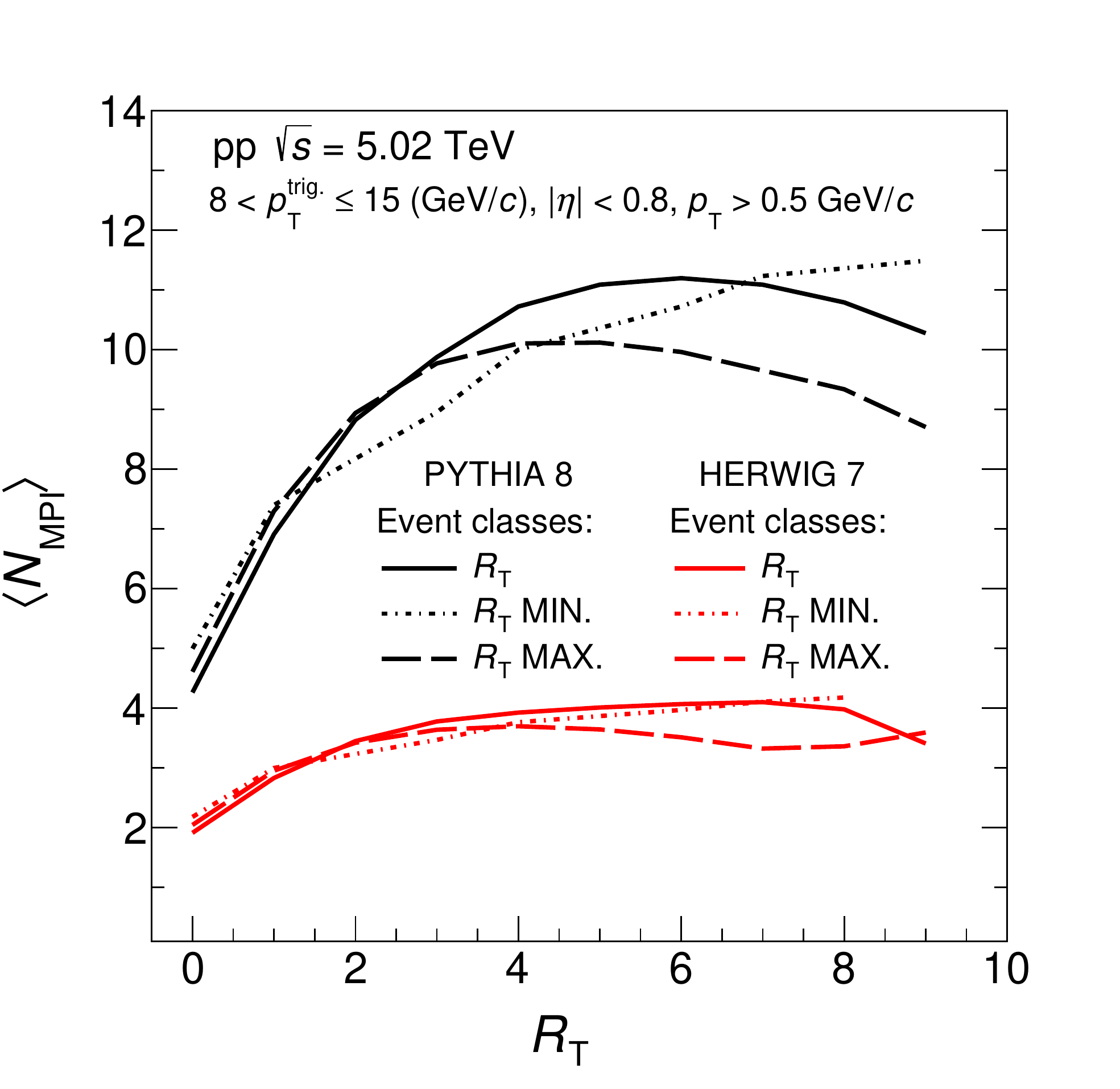}
\includegraphics[width=0.46\textwidth]{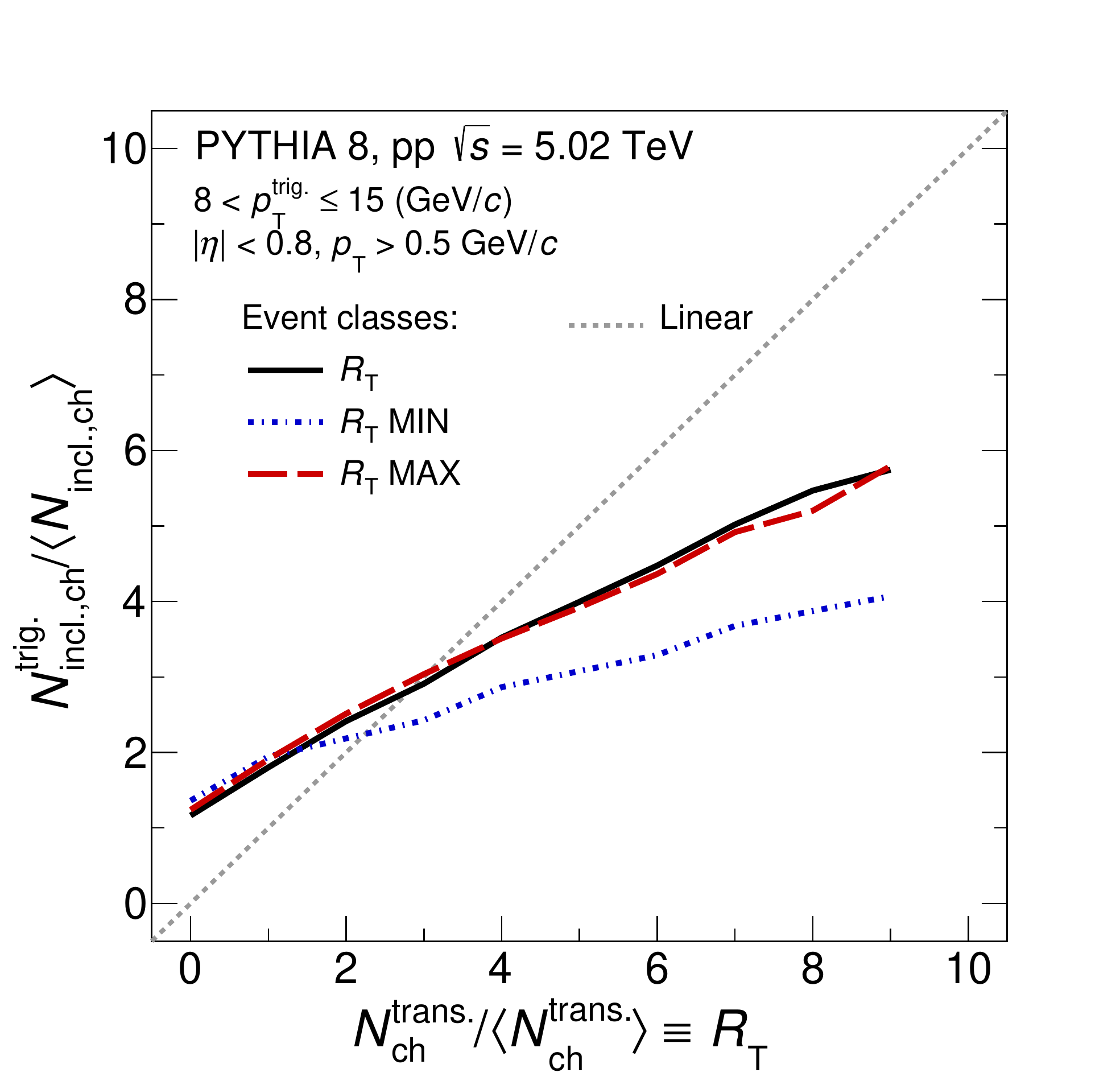}
\caption{Top: Average number of Multiparton Interactions as a function of transverse activity classifiers \rt (dashed line), \rtmin (solid line) and \rtmax (dotted dashed line) for three transverse regions: inclusive, trans-min and trans-max, respectively (see the text for more details). Bottom: \rt dependence of the average charged particle multiplicity in minimum-bias pp collisions ($\pt\geq0.5$\,GeV/$c$ and $|\eta|<0.8$) normalized to that calculated for events which have a trigger particle within $8<\pt^{\rm trig.}\leq 15$\,GeV/$c$.}
\label{fig:nmpiRT}  
\end{figure}
The top panel of Fig.~\ref{fig:nmpiRT} shows the correlation between the average number of MPI, $\langle\nmpi\rangle$, and \rtmin, \rtmax and \rt. In our previous study~\cite{Ortiz:2020dph}, we showed the inclusive case (\rt) and concluded that $\langle\nmpi\rangle$ (after an initial rise) experiences a depletion with \rt when one approaches several times the average $N^{\rm trans}_{\rm ch}$ which is due to the presence of additional jets in the transverse region. Now, when we differentiate between low and high activity regions, one can immediately observe that $\langle\nmpi\rangle$ as a function of \rtmax follows the same evolution as observed in the inclusive case. However, the size of the gluon radiation effect is different, i.e. the reduction of $\langle\nmpi\rangle$ is more pronounced for \rtmax than for \rt. Contrarily, the trans-min region always indicates a moderate increase with \rtmin. For all the three curves the same behavior is observed up to around $\rt=3.5$, while from this value onward an opposite trend is seen for the trans-max and trans-min regions. 

Qualitatively, both \py~8 and \hw~7 exhibit the same effect, but it is clear that \hw produces a smaller number of MPIs. Since the charged-particle multiplicity distributions (shown in Fig.~\ref{fig:fig1}) are compatible (qualitatively and quantitatively) for both generators, this would imply that for a given $\langle\nmpi\rangle$ a larger number of charged particles are produced in \hw than in \py.  This could be explained in terms of the differences in the hadronization models: \py implements the Lund string fragmentation model~\cite{Andersson:1983ia}, whereas in \hw the cluster model~\cite{Webber:1983if} is used~\cite{gieseke_a2021a}.

\begin{figure*}
\centering
\includegraphics[width=2.\columnwidth, keepaspectratio]{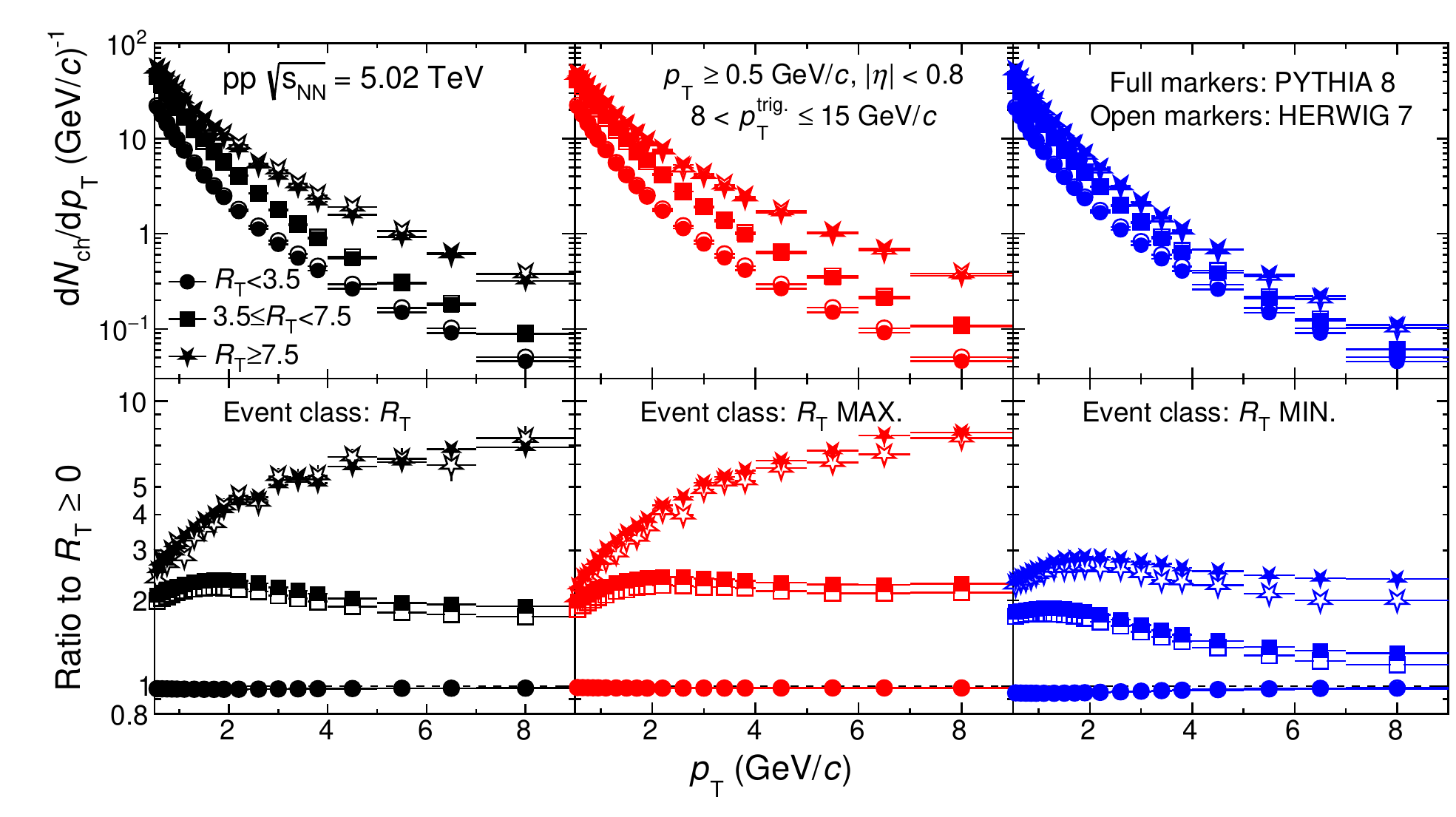}%
\begin{center}
    \caption{Top row: Transverse momentum distributions of charged particles determined for the three topological regions: transverse (left), trans-max (middle) and trans-min (right). The distributions are presented as a function of the event activity (\rt, \rtmax and \rtmin) in the corresponding topological region. Bottom row: The event activity dependent \pt spectra are normalized to the \pt distribution of the \rt-integrated ($\rt>0$) event class. Note the log scale in the $y$ axis.
    \label{fig:ptratio} 
    }
\end{center}
\end{figure*}

It is informative to compare the relative transverse activity classifiers (\rt, \rtmin and \rtmax) to the inclusive average charged-particle multiplicity value $\langle N_{\rm incl,ch}\rangle$, where this average is obtained from events without making any selection on $\pt^{\rm trig.}$. The $\langle N_{\rm incl,ch}\rangle$ is the standard quantity measured in minimum bias collisions by the experiments, however, in this case apart from the selection of all charged particles in $|\eta|<0.8$ a cut $\pt\geq0.5$\,GeV/$c$ is also imposed. The quantity $N^{\rm trig.}_{\rm incl,ch}$ is the total charged particle multiplicity (with the selection $|\eta|<0.8$ and $\pt\geq0.5$\,GeV/$c$) for events having a trigger particle with  $8<\pt^{\rm trig.}\leq15$\,GeV/$c$. The ratio $N^{\rm trig.}_{\rm incl,ch}/\langle N_{\rm incl,ch}\rangle$ is useful to determine our multiplicity reach, and to compare it with that achieved by experiments at the LHC. For example, strangeness enhancement has been observed in pp collisions at $N_{\rm incl,ch} / \langle N_{\rm incl,ch}\rangle \sim 3$~\cite{ALICE:2017jyt}. Based on the Run 2 data from the ALICE Collaboration, the jet quenching searches in pp collisions have been conducted covering multiplicities up to about 5 times the mean multiplicity~\cite{ALICE:2020fuk}. The bottom panel of Fig.~\ref{fig:nmpiRT} shows $N^{\rm trig.}_{\rm incl,ch}/\langle N_{\rm incl,ch}\rangle$ as a function of the relative transverse activity classifier. The slope of each curve is smaller than one below about $\rt=2-3$ reflecting the event selection bias. Moreover, $N^{\rm trig.}_{\rm incl,ch}/\langle N_{\rm incl,ch}\rangle$ exhibits a steeper rise as a function of \rt (and \rtmax) than for \rtmin. For $\rt \approx 8$ ($\rtmax \approx 8$), the inclusive multiplicity reach is about 6 times the mean multiplicity. Whereas, for $\rtmin \approx 8$ it amounts to about 4 times the mean multiplicity. Albeit in this analysis we reached the highest multiplicity values explored so far at the LHC, we are still too far from the extremely large multiplicities which are expected to be collected in the LHC Run 3 by the ALICE Collaboration~\cite{ALICE:2020fuk}. However, with the multiplicity values reached in this analysis, we will be able to understand the effects of hard Bremsstrahlung gluons on the \pt spectra associated to events with large \rt values (up to $\langle N_{\rm MPI} \rangle \sim 12$), and to test our proposed approach to disentangle that effect from the underlying event.

\section{\label{sec:results}Results and Discussion}

In the following studies we will consider at least two distinct \rt intervals: $\rt<3.5$ and $\rt\geq3.5$ which aim to represent the event classes for low and high values of \rt. Furthermore, we will  consider one additional \rt bin between $\rt=3.5$ and $\rt=7.5$, where statistics allows.

First of all, we investigate the transverse momentum distributions of charged particles determined for the three topological regions: transverse, trans-max, and trans-min. Figure ~\ref{fig:ptratio} shows the results for pp collisions at $\sqrt{s}=5.02$\,TeV simulated with both \py~8 and \hw~7 models. The left top panel of Fig.~\ref{fig:ptratio} displays the \pt spectra determined in the transverse regions for three particular \rt bins: $\rt<3.5$, $3.5\leq\rt<7.5$, and $\rt\geq7.5$.  The bottom panel shows the event activity and \pt dependent ratios: \pt spectra normalized to the \pt distribution of the \rt-integrated event class ($\rt\geq0$). Generally one can observe that the larger the \rt the harder the \pt spectrum becomes with respect to the $\rt\geq0$ case; similar observation was made recently by the ALICE Collaboration when both the multiplicity and the  \pt spectra are measured within the same pseudo-rapidity interval $(|\eta|<0.8)$~\cite{Acharya:2019mzb}. It is worth noting that the ALICE preliminary results for the \pt spectra in the transverse side as a function of \rt also follow the same trend~\cite{Tripathy:2021fax}. 

As it is discussed earlier, we can study the event activity as a function of the multiplicity in the trans-max region, which helps to separate the ``hard component'' (initial and final-state radiation) from the softer one within the same pp collision~\cite{Buttar:2005gdq}. The middle panel shows the result for the trans-max region, which indicates a slightly larger high-\pt production than that observed in the transverse side. This is consistent with the concept that \rtmax selects the hardest component of UE. Qualitatively, both ~\py~8 and \hw~7 give the same result. The right panel of Fig.~\ref{fig:ptratio} depicts the evolution of the \pt spectra as a function of \rtmin. Even though both the average number of MPI and the average particle density at $|\eta|<0.8$ are roughly the same for the similar \rt, \rtmin and \rtmax intervals, the behaviour of the ratios is qualitatively different. At high \pt, the ratios become flat for all the \rtmin intervals, whereas at intermediate \pt, the ratios exhibit an overall increase with increasing \rtmin. At the same time, we observe the development of a bump at $\pt\approx 2-3$\,GeV/$c$. This behaviour is qualitatively similar to that seen when the analysis is performed as a function of the number of MPI. Namely, in Ref.~\cite{Ortiz:2020rwg}, a Cronin-like peak is reported in \py events with large number of MPI. The effect is attributed to color reconnection, which is known to produce flow-like behaviour~\cite{Ortiz:2013yxa}. It is worth mentioning that this bump structure has neither been seen in the analysis of unidentified charged particle production as a function of the mid-pseudorapidity charged particle multiplicity nor using the estimator based in the forward VZERO detector of the ALICE Collaboration~\cite{Acharya:2019mzb}. The possible reason for this is that the selection biases hide the MPI and color reconnection effects.

\begin{figure*}
\begin{center}
\includegraphics[width=0.96\textwidth]{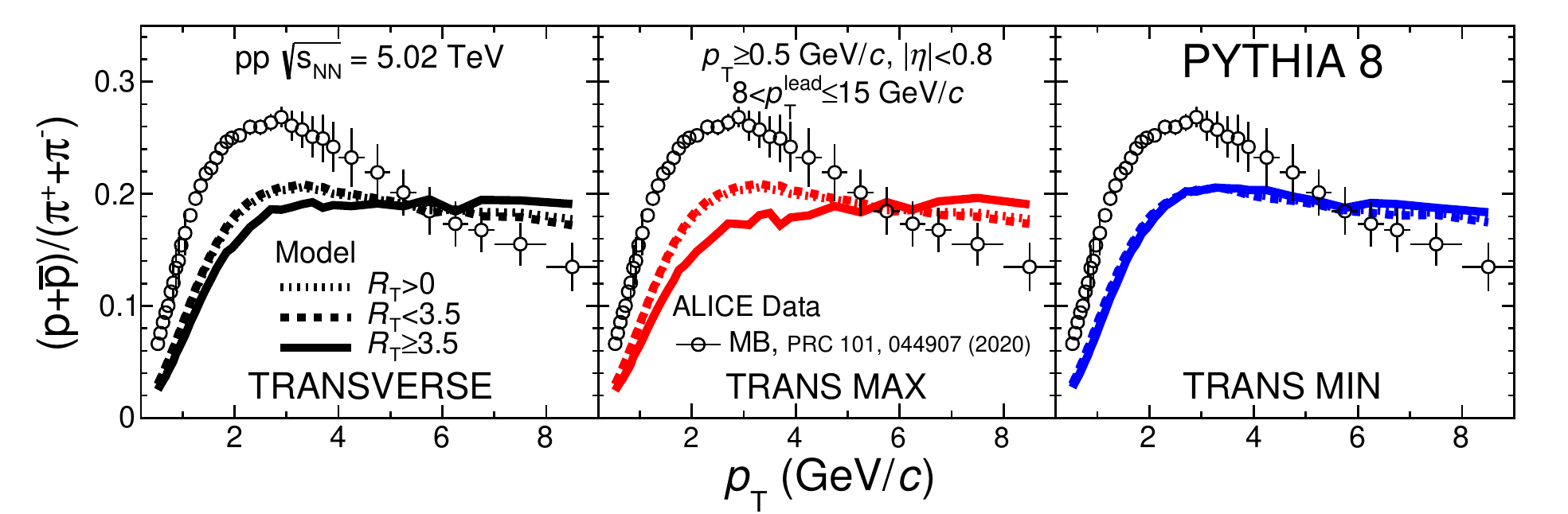}
\includegraphics[width=0.96\textwidth]{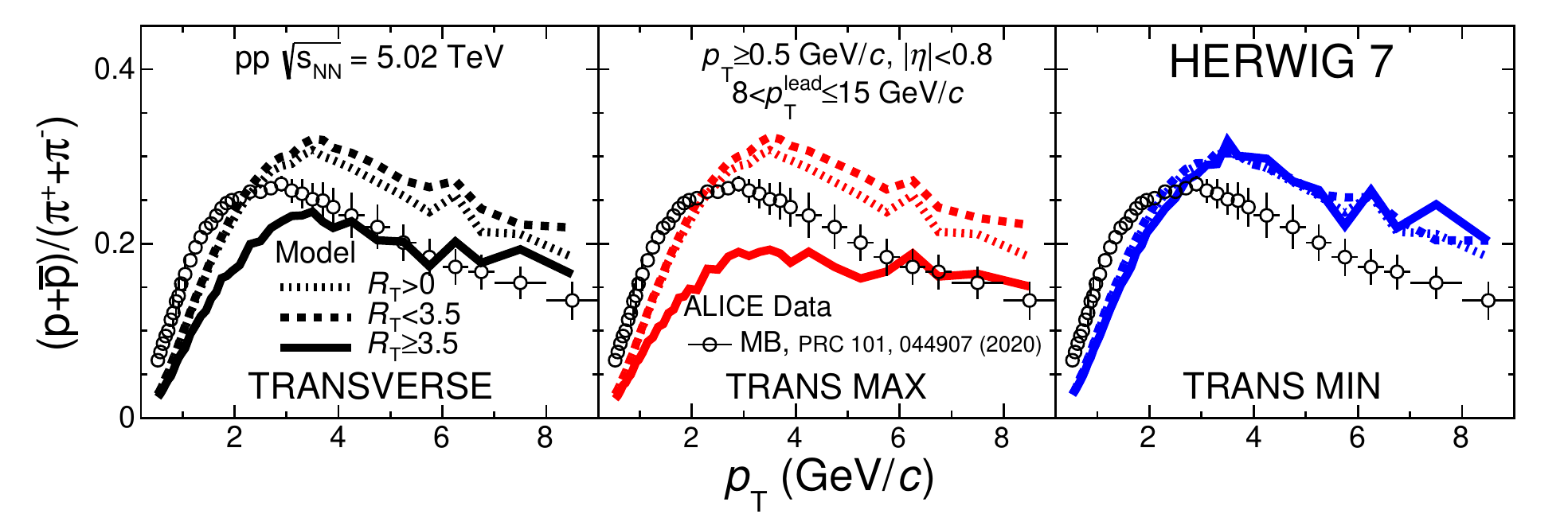}
\caption{Proton-to-pion ratio as a function of \pt for the transverse (left panel), trans-max (middle panel), and trans-min (right panel) regions for pp collisions at $\sqrt{s}=5.02$\,TeV simulated with \py~8 (upper panel) and \hw~7 (bottom panel). For the transverse region, results are shown for low- ($\rt<3.5$), high- ($\rt\geq3.5$), and inclusive ($\rt>0$) \rt selections. For trans-max (trans-min) the analysis is performed as a function of \rtmax (\rtmin) instead of \rt. Model calculations are compared with minimum-bias data (markers) from the ALICE experiment measured at the same $\sqrt{s}$~\cite{Acharya:2019yoi}.}
\label{fig:ptopi}  
\end{center}
\end{figure*}

Apart from charged particle results, we also investigated the \pt-dependent particle ratios of baryons and mesons for transverse, trans-max and trans-min regions. In particular, we calculated the proton-to-pion particle ratios in these topological regions. The left panel of Fig.~\ref{fig:ptopi} shows the p/$\pi$ particle ratios as a function of \pt for the transverse region. Results are shown for low- ($\rt<3.5$), high- ($\rt\geq3.5$), and inclusive ($\rt>0$) transverse activity selections simulated with \py~8 (upper panel) and \hw~7 (bottom panel).  The middle (right) panel of Fig.~\ref{fig:ptopi} shows the analogous analysis for the trans-max (trans-min) region, the ratios are presented as a function of \rtmax (\rtmin).  The model results are compared with minimum bias data from the ALICE experiment measured at the same $\sqrt{s}$~\cite{Acharya:2019yoi}. Experimental results of baryon-to-meson ratios measured in the jet and ``bulk'' (underlying event) regions show that the ratios in jets is below the measured ratio of inclusive particles, and the enhancement of the baryon-to-meson ratio thus seems to be due to bulk effects~\cite{Zimmermann:2015npa}. One would therefore expect to see an enhancement of the ratio with increasing the underlying event activity (\rt). However, both \py~8 and \hw~7 give an opposite behaviour when the event selection is done either using \rt or \rtmax. In order to shed light on this effect we examine further the p/$\pi$ ratio.

For the results obtained with \py~8, we see that the p/$\pi$ ratio for the transverse region (left panel) features a bump structure for the $\rt$-integrated ($\rt>0$) as well as for the low-\rt event class, and its evolution is similar for the trans-max and trans-min regions. However, when a high-\rt class is considered, a significant difference is witnessed between low- and high-\rt cases. Not only does its shape indicates difference with respect to the low-\rt class, the magnitude of the ratio is reduced as well. The middle panel of Fig.~\ref{fig:ptopi} reports the high-activity transverse region which\,---\,as we discussed earlier\,---\,receives large amount of particles from hard gluon-radiated processes, i.e. hard gluons produced in ISR/FSR radiation. In effect the previously observed bump structure is smeared out and a similar effect is seen as for particle ratios inside jets~\cite{Zimmermann:2015npa}. Hard gluons from ISR/FSR can be responsible for this phenomenon because during their large-angle radiation the multiplicity of the event can increase and in turn it might populate the transverse region as well. It is interesting to look at the p/$\pi$ ratio when minimum activity is measured in the transverse region (right panel). As mentioned earlier, the average \nmpi increases with \rt as opposed to what is observed for trans-max. The shape of the p/$\pi$ ratio is similar to what observed for the low-\rt trans-max, however, its \rt evolution is different. Namely, a small increase of the ratio at $\pt \sim 3$\,GeV/$c$ with increasing \rtmin is observed. In other words, the ratio is mostly composed of particles coming from softer parton-parton scatterings which consist of the UE. Similar observations can be made for the \hw~7 model. Few differences are observed: 1) there is a stronger dependence on the chosen \rt intervals, 2) the model better represents the data in terms of their \pt evolution above about $\pt=4$\,GeV/$c$, however these do not interfere with our previous conclusions seen in terms of the event activity dependencies.

\begin{figure*}
\centering
\includegraphics[width=0.66\columnwidth, keepaspectratio]{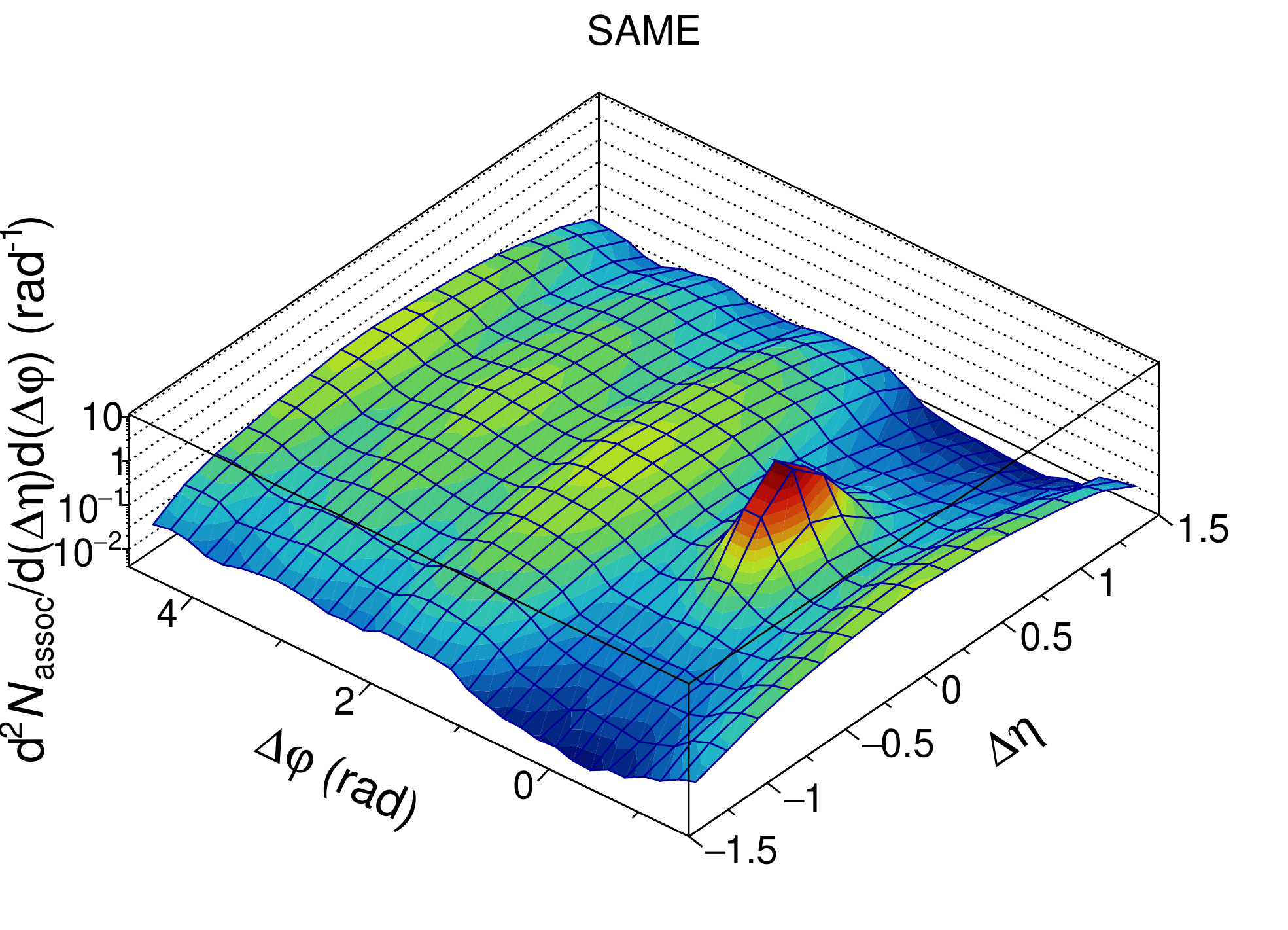}%
\includegraphics[width=0.66\columnwidth, keepaspectratio]{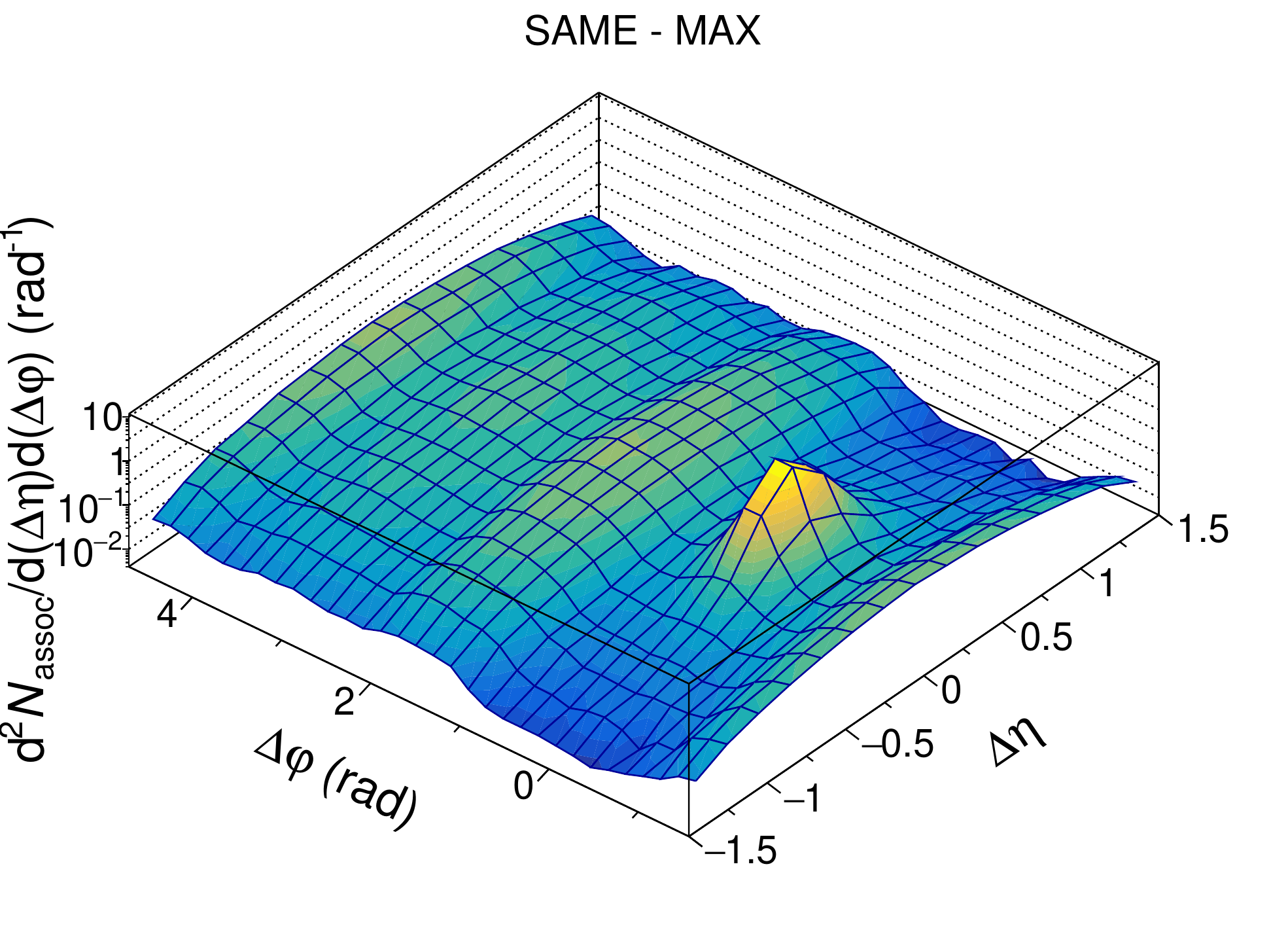}%
\includegraphics[width=0.66\columnwidth, keepaspectratio]{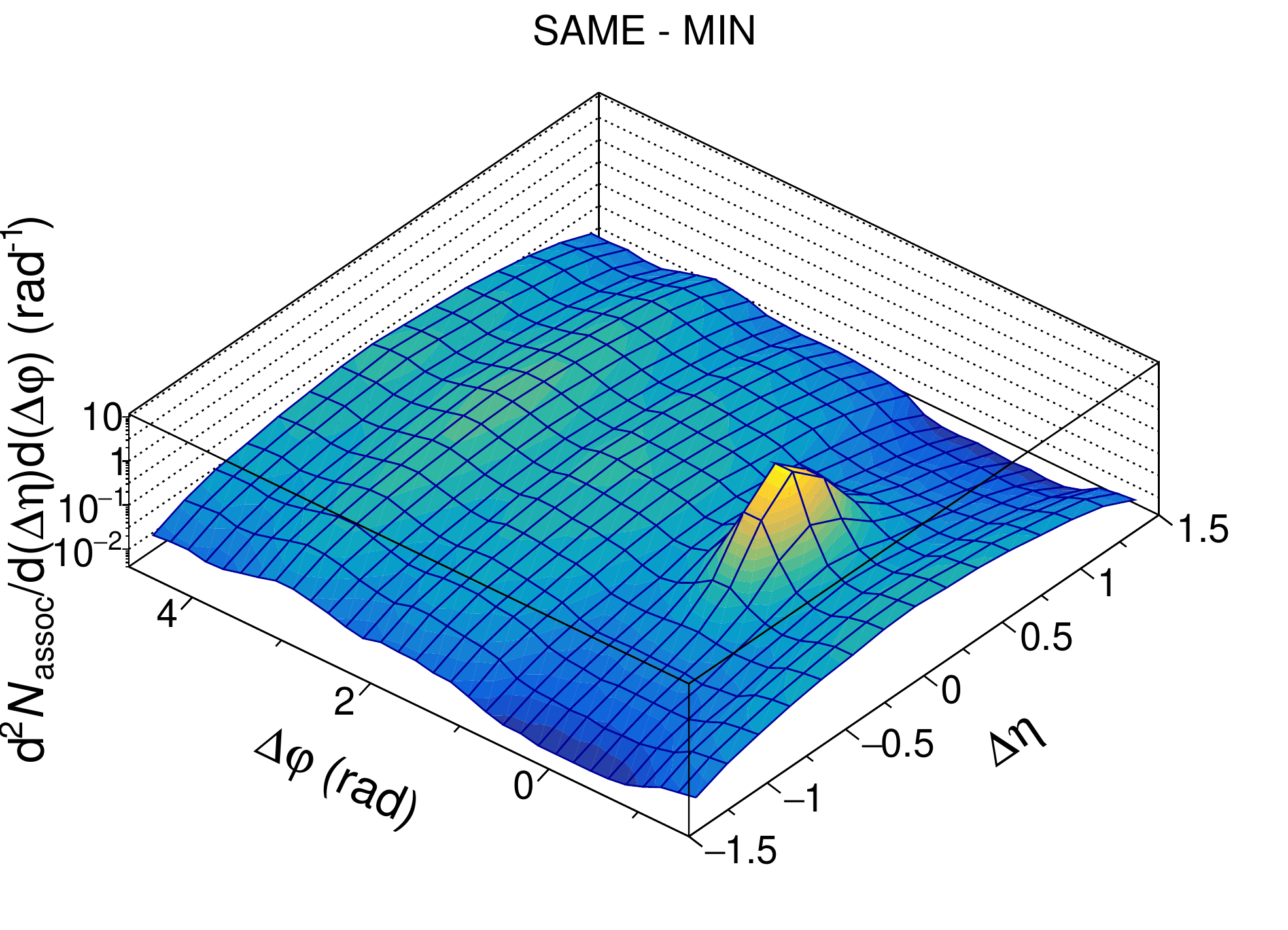}\\%
\includegraphics[width=0.66\columnwidth, keepaspectratio]{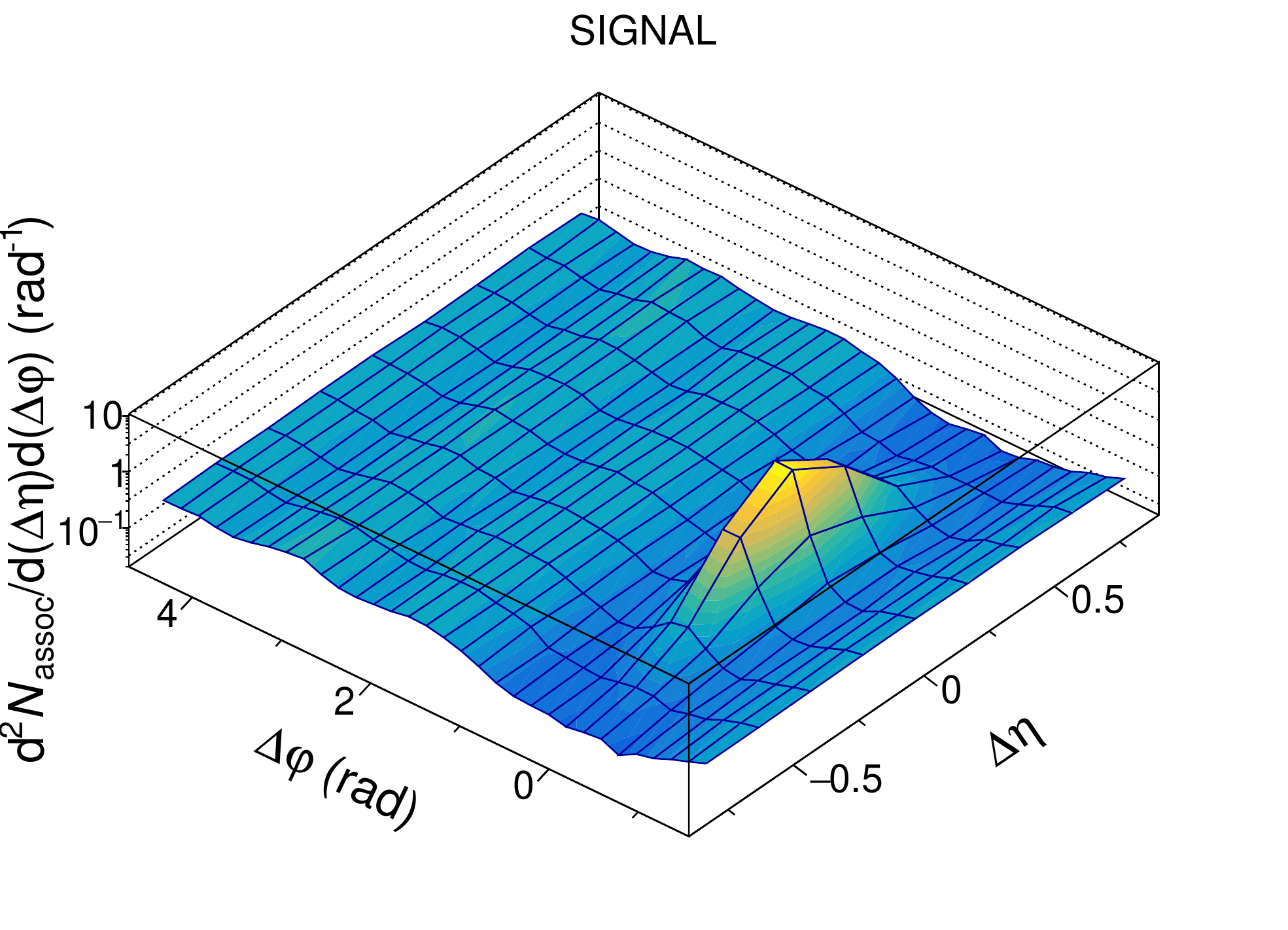}%
\includegraphics[width=0.66\columnwidth, keepaspectratio]{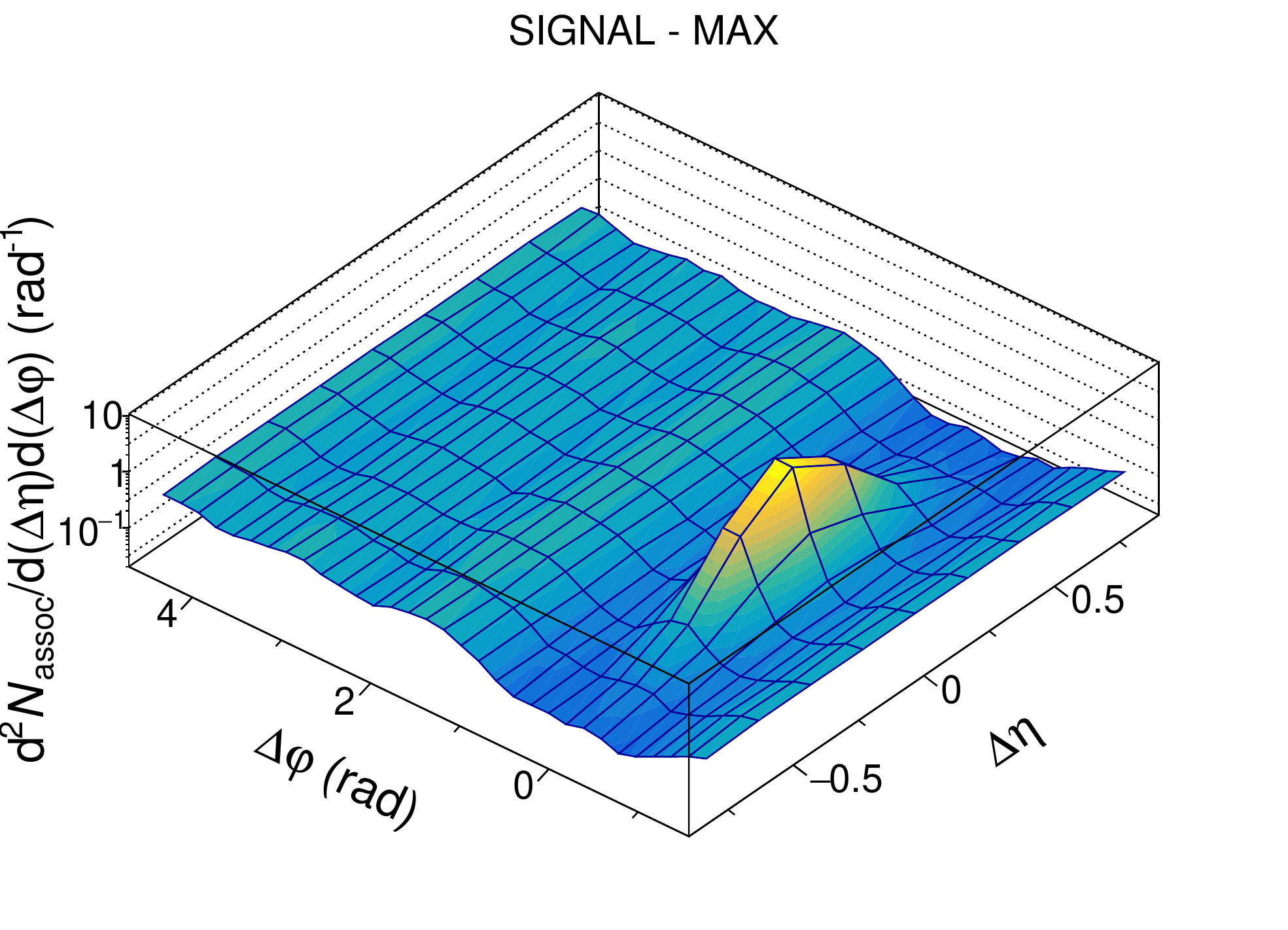}%
\includegraphics[width=0.66\columnwidth, keepaspectratio]{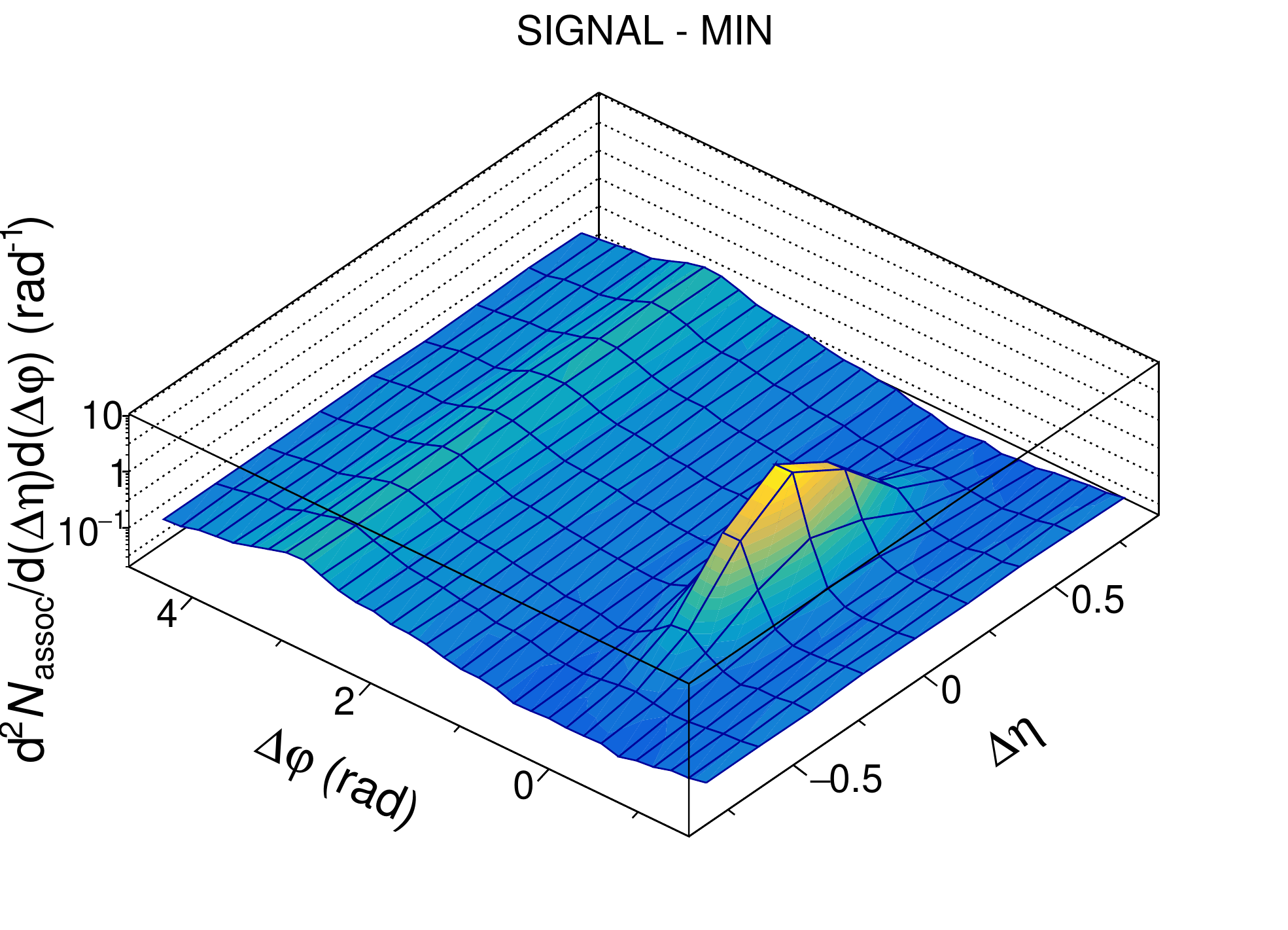}%
    \caption{\label{fig:dphideta} 
    Charged particle yield as a function of $\Delta\eta-\Delta\phi$ for the event class $\rt\geq3.5$ simulated with \py~8 in pp collisions at $\sqrt{s}=5.02$\,TeV. The associated particles have a \pt in the range $4\leq\pt^{assoc.}<6$\,GeV/$c$. Top row shows the distributions for the same event, whereas bottom row shows the signal obtained after removing the mixed event (see the text for details). Second and third columns represent the cases of trans-max and trans-min corresponding to low and high activities in the transverse region. 
    }
\end{figure*}

\begin{figure*}
\centering
\includegraphics[width=\columnwidth, keepaspectratio]{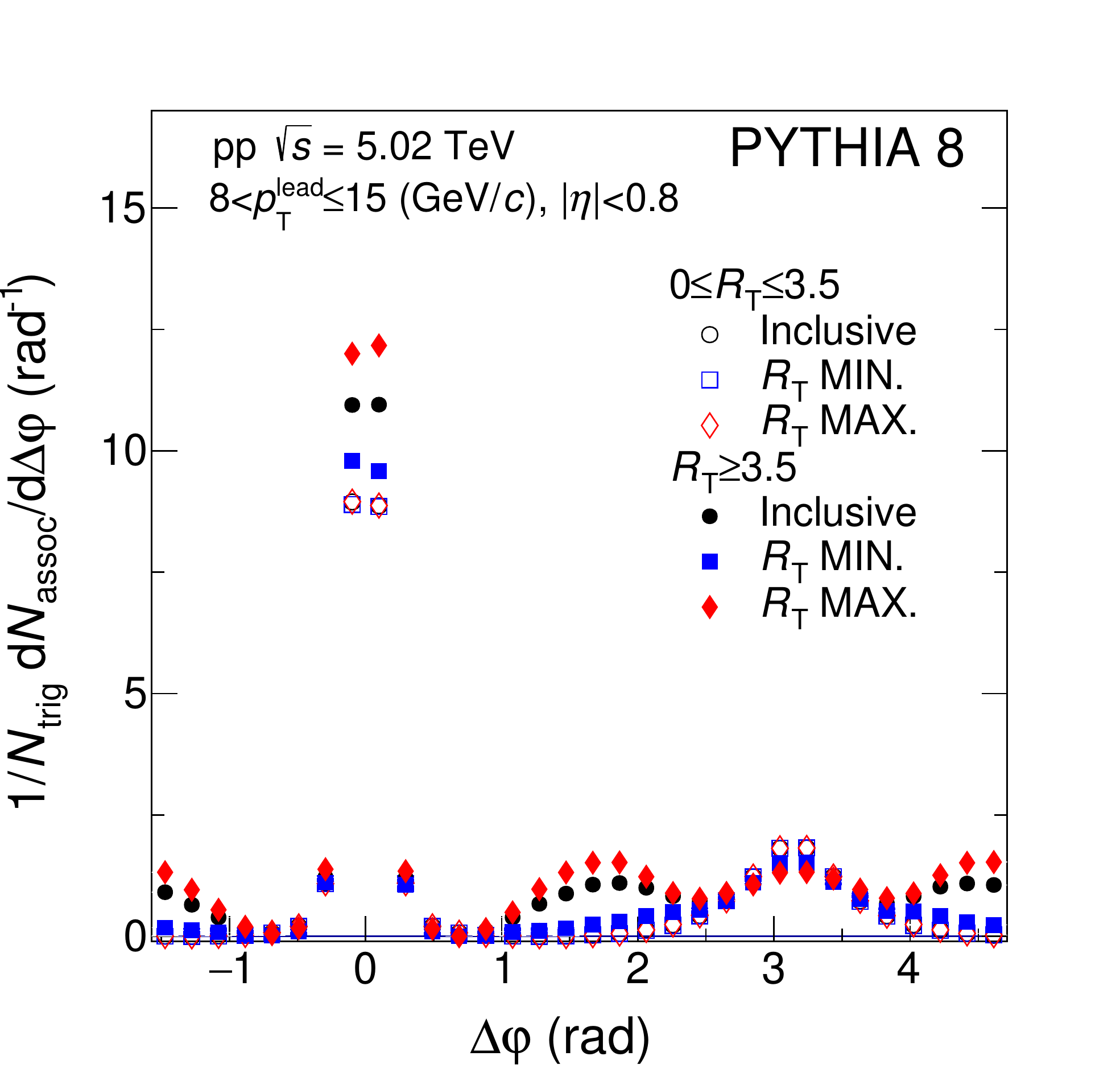}%
\includegraphics[width=\columnwidth, keepaspectratio]{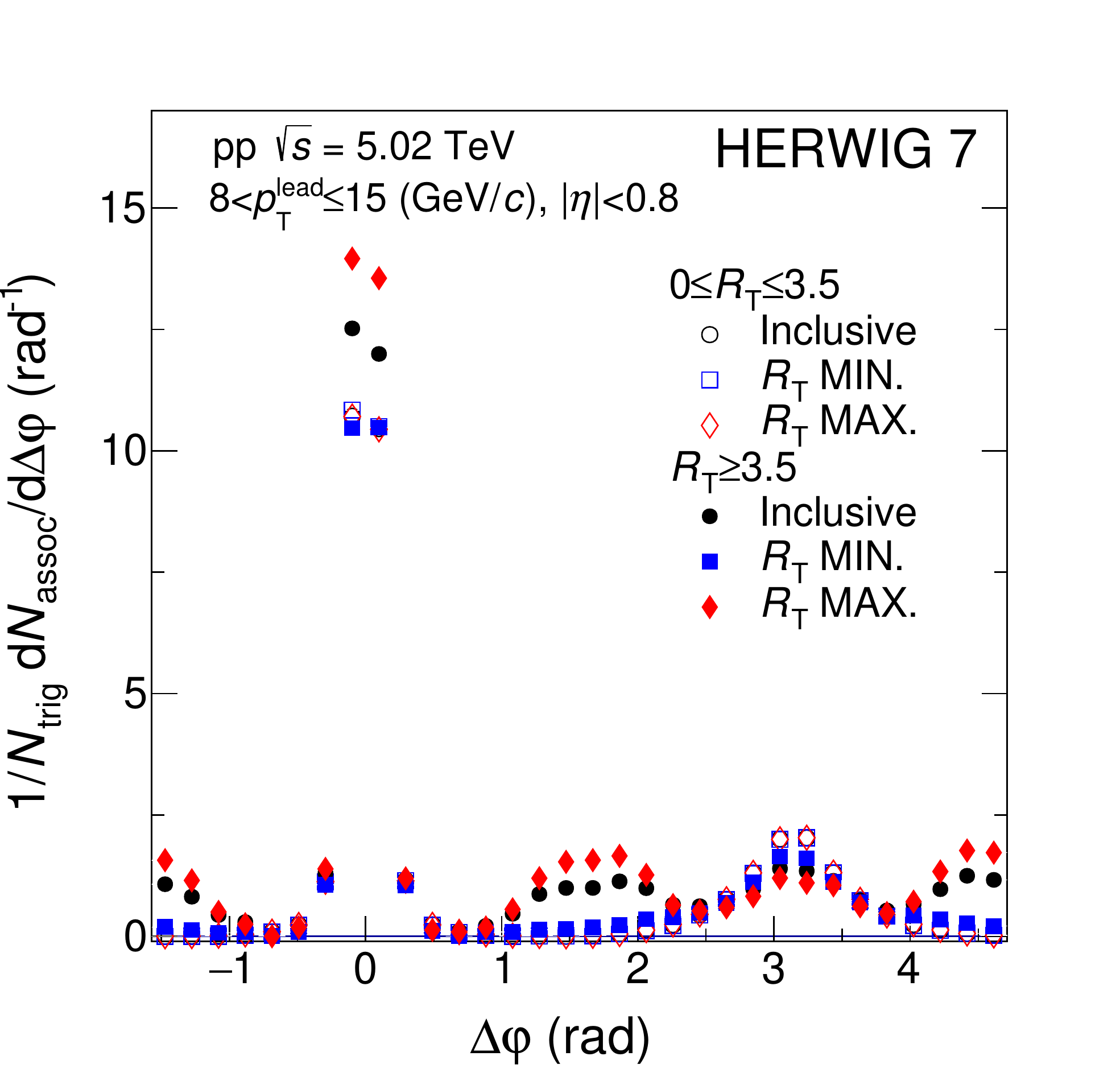}
\begin{center}
    \caption{Charged-particle yield as a function of $\Delta\phi$. Results are shown for \py~8 (left) and \hw~7 (right). The distributions are shown for the transverse, trans-max (\rt max.), and trans-min (\rt min.) regions for low- ($\rt<3.5$), high- ($\rt\geq3.5$) transverse activity selections.
    \label{fig:figdphi} 
    }
\end{center}
\end{figure*}

Last but not least, in our previous study~\cite{Ortiz:2020dph} we showed the di-hadron correlation as a function of $\Delta\eta$ ($= \eta^{\rm trig.}-\eta^{\rm assoc.}$) and $\Delta\phi$ with associated particles having $4\leq\pt^{\rm assoc.}<6$\,GeV/$c$ for low values of \rt. We observed that due to the event selection a structure on the transverse region emerges with the increase of \rt and its shape is well reproduced by the mixed event distribution. Here, we investigate further the structure in the transverse region when we differentiate between low- and high activity transverse regions. Since interesting effects appear between these two activity classes only at higher \rt values, we focus our attention to $\rt\geq3.5$ and compare it to $\rt<3.5$. We remind the reader that in Ref.~\cite{Ortiz:2020dph} we already studied the low-\rt region (although with a finer binning), and we observed that the signal distribution showed no sizable structure in the transverse region; more precisely, the transverse activity starts to be important only at around $\rt=3.5$ and above. 

Figure~\ref{fig:dphideta} shows the per-trigger charged particle yield as a function of $\Delta\eta-\Delta\phi$ for the event class $\rt\geq3.5$. Top row presents the distributions for the same event, whereas bottom row shows the signal which is obtained after the subtraction of the mixed-events correlation. Similarly to our previous work, the mixed events are classified with the charged-particle multiplicity in the transverse region, however, considering now also situations when we make a distinction between the low- and the high-activity transverse regions. Comparing the $\Delta\eta-\Delta\phi$ same distributions for the \rt-inclusive, \rtmin, and \rtmax we can immediately see that the transverse structure is always present in the high-\rt class ($\rt\geq3.5$), although with unequal magnitudes. It is remarkably reduced for \rtmin, and one cannot observe it anymore (with the current normalization) in the signal distribution (bottom right panel). We further investigate UE region by studying the projection of the $\Delta\eta$-$\Delta\phi$ distribution to the $\Delta\phi$ axis. Figure~\ref{fig:figdphi} shows the charged-particle yield as a function of $\Delta\phi$. Left and right panels show the results obtained with \py~8 and \hw~7, and compares the \rt-inclusive, \rtmax), and \rtmin event classes for low- ($\rt<3.5$) and high- ($\rt\geq3.5$) transverse activity selections. While for $\rt<3.5$ there is no quantitative difference between the \rt activity selections, for $\rt\geq3.5$ one can observe a clear evolution of the yields with the \rt activity. The activity in the transverse side is higher in events with large \rtmax than in those having large \rt. This is explained by the presence of increased ISR and FSR in events with large \rtmax. In contrast, using \rtmin instead of \rt or \rtmax, we see a significant reduction of the yields, and the remaining signal in the transverse region is \rtmin independent to a large extent. This effect elucidates the decreasing role of ISR/FSR effects, and in turn the selection bias. It is noteworthy to look at the away region $|\Delta\phi|>2\pi/3$) as well: we observe a broadening of the signal distribution with increasing \rt or \rtmax. When \rtmin is used, there is little or no evolution is seen with increasing \rtmin values.

\section{Conclusions}

In this work we studied the particle production as a function of the underlying-event activity quantified in terms of the relative transverse activity classifier, \rt. The study was performed for pp collisions at $\sqrt{s}=5.02$\,TeV simulated with the Monte Carlo event generators \py~8 and \hw~7. We proposed a modification to the \rt definition of Peter Skands et al.~\cite{Martin:2016igp}. This was motivated by recent preliminary ALICE results which show a strong bias when both the \pt spectra and the event multiplicity are determined in the same transverse region of the di-hadron correlations: $|\Delta\eta|<0.8$ and $\pi/3<|\Delta\phi|<2\pi/3$. The bias is attributed to hard  Bremsstrahlung gluons originated by initial- and final-state radiation~\cite{Ortiz:2020dph}. In order to control the sensitivity to hard processes, we propose to split the transverse region into the so-called trans-min and trans-max regions. They were originally introduced in the underlying event analysis by the CDF Collaboration~\cite{Buttar:2005gdq}. Using the activity in trans-min (trans-max) we defined \rtmin (\rtmax) in the same way as the original relative transverse activity classifier. In order to prove the expected performance of the three event classifiers, we calculated the quantities listed below using kinematic cuts ($\pt>0.5$\,GeV/$c$ and $|\eta|<0.8$) which are accessible to experiments at RHIC and at the LHC.

\begin{itemize}
    \item {\bf Transverse momentum spectra of unidentified charged particles:} Given that trans-min is known to be more sensitive to soft Multiparton Interactions. The \rtmin-dependent \pt spectra normalized to the inclusive \pt distribution ($\rt\geq0$) exhibit a bump at $\pt\approx2-3$\,GeV/$c$ in events with large \rtmin. Whereas at high \pt, this ratio is flat. This behaviour (Cronin-like peak) is seen in Monte Carlo event generators which incorporate MPI~\cite{Ortiz:2020rwg}, but it has never been observed in pp data because the event classifiers used by experiments are strongly affected by selection biases. Contrary, the ratios as a function of \rt or \rtmax exhibits a strong \pt dependence with increasing \rt (or \rtmax).  This effect is consistent with the expectation that trans-max is very sensitive to jets originated from initial and final-state radiation. It is also important mentioning that the ~\py~8 and \hw~7 predictions fully describe the existing preliminary LHC data on \pt spectra as a function of \rt~\cite{Tripathy:2021fax}.
    \item {\bf Baryon-to-meson ratio:} The \pt dependent proton-to-pion ratios as a function of \rt (or \rtmax) exhibits a depletion at $\pt\approx2-3$\,GeV/$c$ with increasing \rt (or \rtmax). The results as a function of \rt agree with the preliminary ALICE data~\cite{Nassirpour:2749160}. The effect is more important when the selection is done in terms of \rtmax instead of \rt.  This depletion is consistent with the presence of jets in the transverse region. Namely, it has been reported both in Monte Carlo~\cite{Cuautle:2015kra} and data~\cite{Zimmermann:2015npa} that the baryon-to-meson ratio is smaller in the jet region than in the underlying-event region. Contrary, the particle ratios as a function of \rtmin exhibit an opposite behaviour. The proton-to-pion ratio at $\pt\approx2-3$\,GeV/$c$ exhibit a small enhancement with increasing \rtmin. This behavior is expected in \py~8 events with large number of MPI~\cite{Ortiz:2013yxa,Ortiz:2020rwg}.     
    \item {\bf Di-hadron correlations:} We have also studied the di-hadron correlations as a function of the underlying event activity. In a previous publication, we reported that the yield in the toward region ($|\Delta\phi|<\pi/3$) increases with increasing \rt. Now, we report that the effect is stronger if one considers \rtmax instead. It is also worth mentioning that the activity in the transverse side is higher in events with high \rtmax than in those with high \rt. This is consistent with the presence of more ISR and FSR in events with large \rtmax. Regarding the away side ($|\Delta\phi|>2\pi/3$), we observe a broadening of the signal with increasing \rt (or \rtmax). When one reduces the bias, i. e. using \rtmin instead of \rt or \rtmax, we do not observe a remarkable evolution of both the toward and away regions with increasing \rtmin. Moreover, the remaining signal in the transverse region is much smaller and roughly \rtmin independent. The results encourage performing a jet quenching search using \rtmin as event activity estimator. 
\end{itemize}
The experimental confirmation of the results discussed above could be useful to understand the similarities of pp data with heavy-ion collisions. In particular, having an event classifier with strong sensitivity to the MPI activity could help to isolate special high-multiplicity pp events originated by several soft MPI. This can help to understand QCD dynamics and its connection with the fluid-like behaviour observed in pp collisions. The use of these observables can also be useful to tune the jet quenching searches in pp collisions which currently are affected by the selection biases~\cite{Jacobs:2020ptj} which \rtmin could significantly reduce.

\begin{acknowledgments}
We acknowledge the technical support of Luciano Diaz and Eduardo Murrieta for the maintenance and operation of the computing farm at ICN-UNAM. Support for this work has been received from CONACyT under the Grants No. A1-S-22917 and CF-2042. G. B. acknowledges the postdoctoral fellowship of CONACyT under the Grant No. A1-S-22917.
\end{acknowledgments}

\bibliography{rt}

\end{document}